\documentclass[9pt,dvips]{article}
\usepackage{eurosym}
\usepackage{amsfonts}
\usepackage{graphicx}
\usepackage{amsmath}
\usepackage{amssymb}
\usepackage{latexsym}
\usepackage{color}

\input{tcilatex}
\begin{document}

\thispagestyle{empty}

\begin{center}
\vspace{1cm}

{\Large \textbf{Demonstration of Monogamy Laws for Gaussian Steering in
Optomechanics}}

\vspace{1.5cm}

\textbf{J. El Qars}$^{a,b,e}${\footnote{%
email: \textsf{j.elqars@gmail.com}}}, \textbf{M. Daoud}$^{c,d}${\footnote{%
email: \textsf{m$_{-}$daoud@hotmail.com}}}, \textbf{R. Ahl Laamara}$^{e}${%
\footnote{%
email: \textsf{ahllaamara@gmail.com}}} and \textbf{N. Habiballah}$^{a,b}${%
\footnote{%
email: \textsf{n.habiballah@uiz.ac.ma}}}

\vspace{0.5cm}

$^{a}$\textit{Faculty of Applied Sciences, Ait-Melloul, Ibn Zohr University,
Morocco }\\[0.5em]
$^{b}$\textit{EPTHE, Department of Physics, Faculty of Sciences, Ibn Zohr
University, Agadir, Morocco }\\[0.5em]
$^{c}$\textit{Abdus Salam International Centre for Theoretical Physics,
Miramare 34151, Trieste, Italy}\\[0.5em]
$^{d}$\textit{Department of Physics, Faculty of Sciences, Ibn Tofail
University, K\'{e}nitra 14000, Morocco}\\[0.5em]
$^{e}$\textit{LPHE-MS, Faculty of Sciences, Mohammed V University, Rabat
10090, Morocco}\\[0.5em]

\vspace{0.9cm}\textbf{Abstract}
\end{center}

Secrecy and security are crucial in communication. So, quantum secret
sharing (QSS) protocol has recently been proposed to distribute a secret
message to a set of parties, where the decoding procedure is forbidden
individually, and a cooperative operation is needed. For this, quantum
steering as an intriguing kind of nonlocality, is proven to be a useful
resource for efficient implementations of the QSS protocol. Here, we study
the distribution of Gaussian steering over an asymmetric three-mode
optomechanical state. We show that a single-mode cannot be jointly steered
by the two others, and further verify the monogamy inequalities of Gaussian
steering. The state at hand displays \textit{genuine tripartite steering}.
Also, we observe one-way steering in the configuration (1vs1)-mode as well
as (1vs2)-mode, hence, we reveal that the asymmetry introduced into the
state we consider, is a necessary condition--but not sufficient--for
reaching one-way steering. As well, we detect one-way steering between two
modes never directly interact. Strikingly, our model exhibits an extreme
level of steering, where two single-mode cannot steer individually the third
mode, while, they can collectively, which is decisive for practical
execution of the QSS protocol. \newline

\section{Introduction}

In his landmark paper \cite{Bell}, Bell has shown that bipartite quantum
nonlocality cannot be equivalent to any local hidden variable theory.
Further, Svetlichny raised an interesting question that is, there could be
tripartite quantum states, in which a genuine tripartite nonlocality cannot
be mimicked by any simple nonlocality that could be shared only between two
parties? \cite{Svetlichny}. Such question can be regarded as a prelude for
understanding the patterns by which quantum nonlocality distributes in
composite systems \cite{Reid2}.

From a phenomenological point of view, quantum nonlocality in mixed states
can be manifested in different forms,i.e., Bell-nonlocality \cite{Bell},
Einstein-Podolski-Rosen (EPR) steering \cite{Wiseman}, entanglement \cite%
{entanglement}, and discord \cite{discord}, where all have been viewed as
valuable resources for quantum communication and computational tasks \cite%
{Chuang}. Interestingly enough--as a fundamental trait of quantum
information theory--such correlations cannot be freely shared across
multiple parties of a composite quantum system, thereby, there exist
limitations in their distribution constrained by the so-called \textit{%
monogamy law} \cite{CKW,Adesso3}.

Within the Gaussian framework, entanglement and quantum discord have already
been proven to be monogamous with respect to the Gaussian R\'{e}nyi-2
entropy \cite{Adesso1}. For quantum steering, Reid was partially answered
the question of monogamy by developing a continuous variable constraint with
restrictions to criteria involving up to second order moments \cite%
{Reid2,Reid1}. Besides this, it has been recently proven that the Gaussian
steering measure proposed in \cite{Kogias1} obeys Coffman-Kundu-Wootters
(CKW)-type monogamy inequalities for all Gaussian states of any number of
modes \cite{Kogias2,Lami}.

In this paper, we \textit{theoretically} study tripartite Gaussian steering
in an optomechanical setup. Let us first briefly recall what is intended by
EPR-steering. According to Schr\"{o}dinger, it is a non-local quantum effect
allowing a remotely preparation of set of quantum states via local
measurements \cite{Schrodinger}. Operationally, quantum steering corresponds
to an entanglement certification by an untrusted party \cite{Wiseman},i.e.,
if two parties Alice and Bob share a quantum state $\hat{\varrho}_{AB}$
which is steerable at least in one direction (say from Alice to Bob), then,
Alice can convince Bob (who does not trust Alice) that the shared state $%
\hat{\varrho}_{AB}$ is entangled via local measurements and classical
communications \cite{Kogias1}.

In the hierarchy of quantum correlations, EPR steering stands between
Bell-nonlocaliy and entanglement. More precisely, from a practical point of
view, a bipartite state $\hat{\varrho}_{\mathcal{XY}}$ exhibiting
Bell-nonlocality is necessary steerable in both directions $\mathcal{X}%
\rightarrow \mathcal{Y}$ and $\mathcal{Y}\rightarrow \mathcal{X}$. While,
quantum steering only in one direction is sufficient to certify that the
state $\hat{\varrho}_{\mathcal{XY}}$ is entangled \cite{Wiseman}.

Unlike entanglement and Bell-nonlocality which are shared symmetrically
between two parties $\mathcal{X}$ and $\mathcal{Y}$, quantum steering is a
directional kind of nonseparable quantum correlations \cite{Wiseman},i.e.,
in some circumstances, a mixed bipartite state $\hat{\varrho}_{\mathcal{XY}}$
may be steerable--say--from $\mathcal{X}$ to $\mathcal{Y}$, but not vice
versa \cite{Kogias1}. So, three different classes of steering can be
distinguished: (\textit{i}) two-way steering, where the state $\hat{\varrho}%
_{\mathcal{XY}}$ is steerable in both directions $\mathcal{X}\rightarrow
\mathcal{Y}$ and $\mathcal{Y}\rightarrow \mathcal{X}$, (\textit{ii}) one-way
steering, for which the state $\hat{\varrho}_{\mathcal{XY}}$ is steerable
only from $\mathcal{X}\rightarrow \mathcal{Y}$ or $\mathcal{Y}\rightarrow
\mathcal{X}$, and (\textit{iii}) no-way steering, where the steerability is
not authorized in any direction. Restricting to Gaussian measurements on
Gaussian states, one-way steering has been conclusively observed \cite%
{Exp,Handchen}. Nevertheless, Gaussian measurements are not sufficient to
reveal fairly the non-classical feature of the considered Gaussian state
\cite{Wollmann}. In other words, there exist some classes of Gaussian states
exhibiting one-way steering under Gaussian measurements, whereas, they will
not be as well using certain non-Gaussian measurements \cite{Wollmann}.
Fortunately, conclusive experimental demonstration of one-way steering free
of assumptions on both the quantum state at hand and the type of performed
measurements has been more recently achieved in Ref. \cite{Farzad}. For the
sake of completeness, recall that using the Schmidt decomposition, a pure
state can always be written in a \textit{symmetric form} by means of local
basis change, thereby, the possibility of observing one-way steering
requires mixed inseparable states \cite{Bowles}.

Since the distribution of quantum correlations between remote parties is a
key procedure underlying various quantum information processing, EPR
steering has been identified as a resource for special tasks of secure
quantum protocols,e.g., one-sided device-independent quantum key
distribution \cite{Branciard}, one-sided device-independent randomness
generation \cite{Skrzypczyk}, one-sided device-independent self-testing of
pure maximally and non-maximally inseparable states \cite{Goswami},
subchannel discrimination \cite{Piani}, secure quantum teleportation \cite%
{zarate}, and quantum secret sharing \cite{Kogias2,Exp,Kogias3}. In
addition, the concept of steering has been straightforwardly related to
fundamental open questions in quantum information theory, where--for
instance--it has been employed to give counterexamples of the Peres
conjecture \cite{Peres}.

Quite importantly, secrecy and security are fundamental in communication. In
this context, the quantum secret sharing (QSS) protocol has recently been
proposed to distribute a secret message between several parties \cite%
{Kogias3}. The central idea in such protocol is that the individual access
to the secret message is forbidden, while, decoding it requires a
cooperative operation \cite{Kogias3}. For this, multipartite EPR-steering is
conjectured to be the main ingredient for efficient implementation of the
QSS protocol \cite{Kogias2}.

We emphasize that multipartite EPR-steering has been recently investigated
in various systems \cite{Exp,Theo1,Theo2,Deng1,Deng2}, unfortunately, not
widely in optomechanics. Here, we give strong evidence suggesting that
optomechanical devices can provide a promising platform for practical
implementation of the QSS protocol as well as quantum one-way tasks.

Our interest in optomechanical systems is motivated by the fact that they
are considered as privileged candidates, in which non-classical effects
predicted by quantum mechanics can be tested \cite{Marquardt}. Proposals
include cooling of a mechanical oscillator near its ground-state \cite%
{cooling}, quantum-state transfer \cite{Qtransfer}, entanglement \cite%
{Entanglements,Paternostro2}, quantum steering \cite{steering}, and also
test of Bell-type inequalities \cite{Marinkovic}. It should be noticed here
that based on the concept of genuine \textit{N}-partite EPR steering and
thanks to violation of the Reid criteria, genuine tripartite steering in
optomechanics has been achieved in Ref. \cite{YXiang} considering the pulsed
regime.

The paper is organized as follows. In Sec. \ref{sec2}, we introduce our
optomechanical system involving two optical modes and a single mechanical
mode. Next, by solving the quantum Langevin equations governing the dynamics
of the system, we obtain its stationary covariance matrix. In Sec. \ref{sec3}%
, we study the distribution of Gaussian steering and verify the steering
monogamy laws over the three considered modes. Also, we investigate one-way
steering in different mode-partitions. Finally, we draw our conclusions in
Sec. \ref{sec4}.

\section{A three-mode optomechanical system \label{sec2}}

\begin{figure}[tbh]
\centerline{\includegraphics[width=9cm]{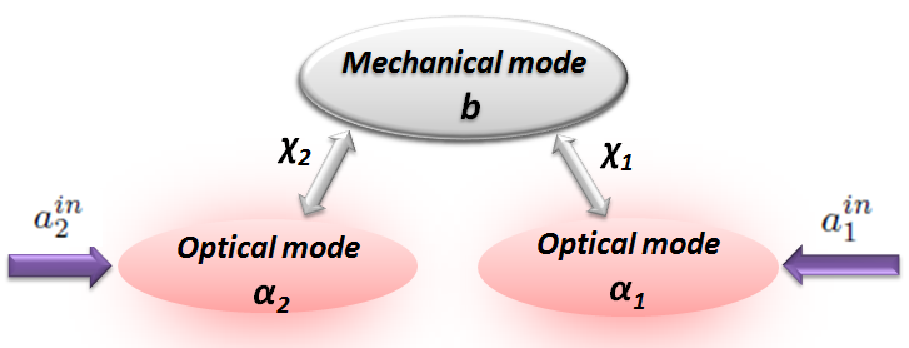}}
\caption{Schematic diagram of the system we consider. The optical mode $a_{j}
$ is coupled to the mechanical one $b$ via the radiation pressure effect
with coupling rate $\protect\chi _{j}$. $a_{j}^{in}$ is the input squeezed
noise affecting the $j\mathrm{th}$ optical mode.}
\label{fig1}
\end{figure}
We consider a cavity system (Fig. \ref{fig1}), with a vibrating mirror
modeled as a single mechanical mode (labelled $C$), with annihilation
operator $b$, an effective mass $\mu $, frequency $\omega _{m}$ and damping
rate $\gamma _{m}$. The mode $C$ is coupled--via the radiation pressure
interaction--to the right(left) optical cavity mode $A$($B$) with
annihilation operator $a_{1}$($a_{2}$), frequency $\omega _{c_{1}}$($\omega
_{c_{2}}$) and decay rate $\kappa _{_{1}}$($\kappa _{_{2}}$). The $j\mathrm{%
th}$ optical mode is driven by a laser source of power $\wp _{j}$, phase $%
\varphi _{j}$ and frequency $\omega _{L_{j}}$. Furthermore, we assume that
the system is fed by two-mode squeezed light, where the first(second) mode
is injected into the right(left) cavity. The dynamics of this three-mode
optomechanical system can be described by the Hamiltonian \cite{Law} $%
\mathcal{\hat{H}}=\hat{\mathcal{H}}_{m}\mathcal{+}\sum%
\limits_{j=1}^{2}\Big(\mathcal{\hat{H}}_{c_{j}}\mathcal{+\hat{H}}_{cp_{j}}%
\mathcal{+\hat{H}}_{in_{j}}\Big)$, where $\mathcal{\hat{H}}_{m}=%
\hbar\omega _{m}b^{\dag }b$($\mathcal{\hat{H}}_{c_{j}}=%
\hbar\omega _{c_{j}}a_{j}^{\dag }a_{j}$) is the free
Hamiltonian of the movable mirror(the $j\mathrm{th}$ cavity mode), and $%
\mathcal{\hat{H}}_{cp_{j}}=(-1)^{j}\hbar\chi
_{j}a_{j}^{\dag }a_{j}(b^{\dag }+b)$ is the optomechanical coupling between
the mechanical mode and the $j\mathrm{th}$ optical mode with coupling rate $%
\chi _{j}=\left( \omega _{c_{j}}/l_{j}\right) \sqrt{\hbar /\mu \omega _{m}}$%
. The last term $\mathcal{\hat{H}}_{in_{j}}=\hbar
\varepsilon _{j}(a_{j}^{\dag }e^{i\varphi _{j}}e^{-i\omega
_{L_{j}}}+a_{j}e^{-i\varphi _{j}}e^{i\omega _{L_{j}}})$ describes the
coupling between the $j\mathrm{th}$ input laser and its corresponding cavity
field with strength coupling $\varepsilon _{j}=\sqrt{2\kappa _{_{j}}\wp
_{j}/\hbar \omega _{L_{j}}}$ \cite{Genes}. Since the system at hand is open,
consequently, it is inevitably affected by dissipation and noise. Thus, the
equations of motion can be obtained using the quantum Langevin equations
(QLEs). So, in a frame rotating with $\omega _{L_{j}}$, we get
\begin{eqnarray}
\partial _{t}b &=&-\bigg(\frac{\gamma _{m}}{2}+i\omega _{m}\bigg)%
b+\sum_{j=1}^{2}(-1)^{j+1}i\chi _{j}a_{j}^{\dag }a_{j}+\sqrt{\gamma _{m}}%
b^{in},  \label{b_dot} \\
\partial _{t}a_{j} &=&-\bigg(\frac{\kappa _{j}}{2}-i\Delta _{j}\bigg)%
a_{j}+(-1)^{j+1}i\chi _{j}a_{j}\left( b^{\dag }+b\right) -i\varepsilon
_{j}e^{i\varphi _{j}}+\sqrt{\kappa _{j}}a_{j}^{in}\text{ \ for \ \ }j=1,2
\label{c_dot}
\end{eqnarray}%
where $\Delta _{j}=\omega _{L_{j}}-\omega _{c_{j}}$ is the $j\mathrm{th}$
laser-cavity detuning \cite{Marquardt}. In Eq. (\ref{b_dot}), $b^{in}$ is
the zero-mean Brownian noise operator acting on the mechanical mode. It is
not $\delta $-correlated in general, however, mirror with large mechanical
quality factor $\mathcal{Q}_{m}=\omega _{m}/\gamma _{_{m}}\gg 1$ allows
recovering the Markovian process,i.e., $\left( \langle b^{in\dag
}(t)b^{in}(t^{\prime })\rangle ;\langle b^{in}(t)b^{in\dag }(t^{\prime
})\rangle \right) =(\bar{n};\bar{n}+1)\delta (t-t^{\prime })$, where $\bar{n}%
=(e^{\hbar \omega _{m}/k_{B}T}-1)^{-1}$ is the mean phonons number \cite%
{Benguria}. $T$ and $k_{B}$ are the temperature of the mechanical bath and
the Boltzmann constant. In Eq. (\ref{c_dot}), $a_{j}^{in}$ is the input
squeezed noise operator (with $\langle a_{j}^{in}\rangle $) affecting the $j%
\mathrm{th}$ optical mode. It is correlated as $(\langle \delta
a_{j}^{in^{\dag }}(t)\delta a_{j}^{in}(t^{\prime })\rangle ;\langle \delta
a_{j}^{in}(t)\delta a_{j}^{in^{\dag }}(t^{\prime })\rangle ;\langle \delta
a_{j}^{in}(t)\delta a_{j^{\prime }}^{in}(t^{\prime })\rangle ;\langle \delta
a_{j}^{in^{\dag }}(t)\delta a_{j^{\prime }}^{in^{\dag }}(t^{\prime })\rangle
)=\mathcal{R}(t,t^{\prime })\delta (t-t^{\prime })$ for $j\neq j^{\prime
}=1,2$, where $\mathcal{R}(t,t^{\prime })=(N;N+1;Me^{-i\omega _{m}\left(
t+t^{\prime }\right) };Me^{i\omega _{m}\left( t+t^{\prime }\right) })$ and $%
N=\mathrm{sinh}^{\mathrm{2}}r$, $M=\mathrm{sinh}r\mathrm{cosh}r$, $r$ being
the squeezing parameter \cite{Paternostro1}. It should be noticed here that
optimal transfer of quantum fluctuations from the squeezed light to the
system can be achieved when the squeezing frequency is resonant with those
of the optical modes,i.e., $\omega _{s}=\omega _{c_{j}}$ \cite{Parkins}.

Due to the nonlinear terms $a_{j}^{\dag }a_{j}$ and $a_{j}\left( b^{\dag
}+b\right) $, Eqs. (\ref{b_dot})-(\ref{c_dot}) are of nontrivial solutions.
However, pumping the system by intense lasers allows the linearization of
the cavities' and mirror' operators around their steady-states \cite{Vitali}%
,i.e., $\mathcal{O}=\langle \mathcal{O}\rangle +\delta
\mathcal{O}$ ($\mathcal{O}\equiv b,a_{j}$), where $\langle
\mathcal{O}\rangle $ and $\delta \mathcal{O}$ are respectively the steady-state mean value (c-number) and a small
fluctuation with zero mean value ($\langle \delta \mathcal{O}\rangle =0$) of
the operator $\mathcal{O}$. Using the fact that $\partial _{t}\langle\mathcal{O}
\rangle=0$ and factorizing the averages in Eqs. (\ref{b_dot}) and (\ref{c_dot}),
hence, we obtain $\langle b\rangle =\sum\limits_{j=1}^{2}\frac{(-1)^{j+1}\chi _{j}\left\vert \langle a_{j}\rangle \right\vert ^{2}}{\omega
_{m}-i\frac{\gamma _{m}}{2}}$ and $\langle a_{j}\rangle =\frac{2\varepsilon
_{j}e^{i\varphi _{j}}}{2\Delta _{j}^{\prime }+i\kappa _{j}}$, where $\Delta
_{j}^{\prime }$ $=$ $\Delta _{j}$ $+(-1)^{j+1}\chi _{j}(\langle b\rangle
^{\ast }+\langle b\rangle )$ is the $j\mathrm{th}$ effective cavity-laser
detuning \cite{Marquardt}. Also, with the above assumption of intense
laser-driven, the nonlinear terms (e.g., $\delta a_{j}^{\dag }\delta a_{j}$)
can be neglected \cite{Genes,Vitali}. In addition, choosing $\tan \varphi
_{j}=-2\Delta _{j}^{\prime }/\kappa _{j}$, we have $\langle a_{j}\rangle
=-i\left\vert \langle a_{j}\rangle \right\vert $, which leads to
\begin{eqnarray}
\ \text{\ \ }\ \text{ \ }\partial _{t}\delta b &=&-\left( \frac{\gamma _{m}}{%
2}+i\omega _{m}\right) \delta b+\sum_{j=1}^{2}(-1)^{j}\bar{\chi}%
_{_{j}}\left( \delta a_{j}-\delta a_{j}^{\dag }\right) +\sqrt{\gamma _{m}}%
b^{in},  \label{delta_b_dot} \\
\partial _{t}\delta a_{j} &=&-\left( \frac{\kappa _{j}}{2}-i\Delta
_{j}^{\prime }\right) \delta a_{j}+(-1)^{j+1}\bar{\chi}_{_{j}}\left( \delta
b^{\dag }+\delta b\right) \ +\sqrt{\kappa _{j}}\delta a_{j}^{in},\
\label{delta_cj_dot}
\end{eqnarray}%
where $\bar{\chi}_{_{j}}$ $=\chi _{j}\left\vert \langle a_{j}\rangle
\right\vert $ is the $j\mathrm{th}$ effective coupling in the linearized
regime \cite{Marquardt}. For simplicity, we introduce the operators $\delta
\tilde{b}=\delta be^{i\omega _{m}t}$ and $\delta \tilde{a}_{j}=\delta
a_{j}e^{-i\Delta _{j}^{\prime }t}$, and consider that the two cavities are
driven in \textit{the red sideband},i.e.,$\Delta _{j}^{\prime }=-\omega _{m}$%
, which is relevant for quantum-state transfer \cite{DWang}. We finally
suppose that the mechanical frequency $\omega _{m}$ is larger than the $j%
\mathrm{th}$ decay rate $\kappa _{j}$ ($\omega _{m}\gg \kappa _{1,2}$) \cite%
{Clerk}, which corresponds to the resolved-sideband regime, so that, the
rotating wave approximation (RWA) allows us to drop the fast oscillating
terms with $\pm 2\omega _{m}$ \cite{Clerk}. Therefore, one gets
\begin{eqnarray}
\partial _{t}\delta \tilde{b} &=&-\frac{\gamma _{m}}{2}\delta \tilde{b}%
+\sum_{j=1}^{2}(-1)^{j}\bar{\chi}_{_{j}}\delta \tilde{a}_{j}+\sqrt{\gamma
_{m}}\tilde{b}^{in},  \label{RWA_b} \\
\partial _{t}\delta \tilde{a}_{j} &=&-\frac{\kappa _{j}}{2}\delta \tilde{a}%
_{j}+(-1)^{j+1}\bar{\chi}_{_{j}}\delta \tilde{b}\ +\sqrt{\kappa _{j}}\delta
\tilde{a}_{j}^{in}  \label{RWA_c}
\end{eqnarray}%
Next, combining the Eqs. (\ref{RWA_b})-(\ref{RWA_c}) with the quadratures
position and momentum of the $j\mathrm{th}$ optical(mechanical) mode,i.e., $%
\delta \tilde{x}_{a_{j}}=(\delta \tilde{a}_{j}^{\dag }+\delta \tilde{a}_{j})/%
\sqrt{2}$ and $\delta \tilde{y}_{a_{j}}=i(\delta \tilde{a}_{j}^{\dag
}-\delta \tilde{a}_{j})/\sqrt{2}$($\delta \tilde{x}_{b}=(\delta \tilde{b}%
^{\dag }+\delta \tilde{b})/\sqrt{2}$ and $\delta \tilde{y}_{b}=i(\delta
\tilde{b}^{\dag }-\delta \tilde{b})/\sqrt{2}$) and their corresponding input
noise operators $\delta \tilde{x}_{a_{j}}^{in}=(\delta \tilde{a}_{j}^{in\dag
}+\delta \tilde{a}_{j}^{in})/\sqrt{2}$ and $\delta \tilde{y}%
_{a_{j}}^{in}=i(\delta \tilde{a}_{j}^{in\dag }-\delta \tilde{a}_{j}^{in})/%
\sqrt{2}$($\delta \tilde{x}_{b}^{in}=(\tilde{b}_{j}^{in\dagger }+\tilde{b}%
_{j}^{in})/\sqrt{2}$ and $\delta \tilde{y}_{b}^{in}=i(\tilde{b}^{in\dagger }-%
\tilde{b}^{in})/\sqrt{2}$), the QLEs can be cast compactly as $\partial _{t}%
\mathcal{\tilde{U}}=\mathcal{K\tilde{U}}+\tilde{\Im}$, where $\mathcal{%
\tilde{U}}^{\mathrm{T}}=(\delta \tilde{x}_{a_{1}},\delta \tilde{y}%
_{a_{1}},\delta \tilde{x}_{a_{2}},\delta \tilde{y}_{a_{2}},\delta \tilde{x}%
_{b},\delta \tilde{y}_{b})$ and $\tilde{\Im}^{\mathrm{T}}=(\delta \tilde{x}%
_{a_{1}}^{in},\delta \tilde{y}_{a_{1}}^{in},\delta \tilde{x}%
_{a_{2}}^{in},\delta \tilde{y}_{a_{2}}^{in},\delta \tilde{x}_{b}^{in},\delta
\tilde{y}_{b}^{in})$. We assume identical decays rates ($\kappa
_{_{1,2}}=\kappa $), and introducing both the damping ratio $\alpha =\gamma
_{m}/\kappa $ \cite{Clerk} and the $j\mathrm{th}$ optomechanical
cooperativity $\mathcal{C}_{j}=4\bar{\chi}_{_{j}}^{2}/\gamma _{m}\kappa =%
\frac{8\omega _{c_{j}}^{2}\wp _{j}}{\gamma _{m}\mu \omega _{m}\omega
_{L_{j}}l_{j}^{2}\left[ \left( \kappa /2\right) ^{2}+\omega _{_{m}}^{2}%
\right] }$ \cite{Purdy}, the kernel $\mathcal{K}$ can be written as $%
\mathcal{K=}\left( \QATOP{\mathcal{K}_{\kappa }}{-\mathcal{K}_{\mathcal{C}}^{%
\mathrm{T}}}\QATOP{\mathcal{K}_{\mathcal{C}}}{\mathcal{K}_{\alpha }}\right) $%
, where $\mathcal{K}_{\kappa }=\frac{-\kappa }{2}%
\mbox{$1
\hspace{-1.0mm}  {\bf l}$}_{4}$, $\mathcal{K}_{\alpha }=\frac{-\alpha \kappa
}{2}\mbox{$1 \hspace{-1.0mm}  {\bf l}$}_{2}$ and $-\mathcal{K}_{\mathcal{C}%
}^{\mathrm{T}}=\left( \QATOP{\frac{-\kappa \sqrt{\alpha \mathcal{C}_{1}}}{2}%
}{0}\QATOP{0}{\frac{-\kappa \sqrt{\alpha \mathcal{C}_{1}}}{2}}\QATOP{\frac{%
\kappa \sqrt{\alpha \mathcal{C}_{2}}}{2}}{0}\QATOP{0}{\frac{\kappa \sqrt{%
\alpha \mathcal{C}_{2}}}{2}}\right) \allowbreak .$

\subsection{ Three-mode covariance matrix}

Since the Eqs. (\ref{RWA_b})-(\ref{RWA_c}) are linearized, $b^{in}$ and $%
a_{j}^{in}$ are zero-mean quantum Gaussian noises, hence, the steady-state
of the quantum fluctuations is a zero-mean three-mode Gaussian state $\hat{%
\varrho}_{ABC}$, where its corresponding $6\times 6$ covariance matrix (CM) $%
\sigma _{ABC}$ is given by $\sigma _{kk^{\prime }}=\langle \{\mathcal{\tilde{%
U}}_{k}(\infty ),\mathcal{\tilde{U}}_{k^{\prime }}(\infty )\}\rangle /2$
\cite{Genes}.\newline
In the Markovian limit, the CM $\sigma _{ABC}$ can be obtained by solving
the Lyapunov equation $\mathcal{K}\sigma +\sigma \mathcal{K}^{\mathrm{T}}=-%
\mathcal{N}$ \cite{Genes,Parks}, where the noise correlations matrix $%
\mathcal{N}$ defined by $\mathcal{N}_{jj^{\prime }}\delta (t-t^{\prime
})=\langle \{\tilde{\Im}_{j}(t),\tilde{\Im}_{j^{\prime }}(t^{\prime
})\}\rangle /2$ can be written as $\mathcal{N}=\mathcal{N}_{\kappa }\oplus
\mathcal{N}_{\alpha }$, with $\mathcal{N}_{\kappa }=\left( \QATOP{\mathcal{N}%
_{1}}{\mathcal{N}_{3}^{\mathrm{T}}}\QATOP{\mathcal{N}_{3}}{\mathcal{N}_{2}}%
\right) $, wherein $\mathcal{N}_{1}\equiv \mathcal{N}_{2}=\frac{\kappa }{2}%
\sinh (2r)\mbox{$1
\hspace{-1.0mm}  {\bf l}$}_{2}$, $\mathcal{N}_{3}=\frac{\kappa }{2}\cosh (2r)%
\mathrm{diag}(1,-1)$, and $\mathcal{N}_{\alpha }=\frac{\alpha \kappa }{2}(2%
\bar{n}+1)\mbox{$1 \hspace{-1.0mm}
{\bf l}$}_{2}.$ It should be noticed that the explicit expression of the CM $%
\sigma _{ABC}$ is too cumbersome to be reported here. However, it can be
simply expressed as
\begin{equation}
\sigma _{ABC}=\left(
\begin{array}{ccc}
\mathcal{M}_{A} & \mathcal{M}_{AB{}}\  & \mathcal{M}_{AC}\  \\
\mathcal{M}_{AB}^{\mathrm{T}}\  & \mathcal{M}_{B}\  & \mathcal{M}_{BC}\  \\
\mathcal{M}_{AC}^{\mathrm{T}}\  & \mathcal{M}_{BC}^{\mathrm{T}}\  & \mathcal{%
M}_{C}%
\end{array}%
\right) ,  \label{CM}
\end{equation}%
where the $2\times 2$ block matrix $\mathcal{M}_{A}(\mathcal{M}_{B})$
represents the first(second) cavity mode $A$($B$), while, $\mathcal{M}_{C}$
represents the mechanical mode $C$. The correlations between the different
modes are given by the $2\times 2$ blocks matrices $\mathcal{M}_{jj^{\prime
}}$ ($j\neq j^{\prime }\in \{A,B,C\}$). As long as the system is driven in
the red sideband, the stability conditions have been verified to be always
satisfied in the chosen parameter regime \cite{Genes}.

\section{Gaussian steerability and its monogamy\label{sec3}}

Classical correlations can be freely shared over a set of parties,i.e., one
party can simultaneously share maximum classical correlations with the
remaining parties \cite{Dhar}. In contrast, there exists a constraint
governing the shareability of quantum correlations across multipartite
states. Such limitation called \textit{monogamy law}, has been originally
expressed by Coffman, Kundu and Wootters (CKW), as a quantitative inequality
traducing the behavior of entanglement among three-qubit states,i.e., $%
\mathfrak{C}_{A/(BC)}^{2}\geqslant \mathfrak{C}_{A/B}^{2}+\mathfrak{C}%
_{A/C}^{2}$, where $\mathfrak{C}_{A/(BC)}^{2}$ is the squared concurrence quantifying entanglement in the bipartition $A/(BC)$ \cite{CKW}. On Gaussian
states, analogous formulas to the qubit case are obtained using the
logarithmic negativity \cite{Illuminati}, and the Gaussian R\'{e}nyi-2
entropy \cite{Adesso1,Lami} as quantifiers of entanglement. On the other
hand, with respect to the measure proposed in Ref. \cite{Kogias1}, it has
recently been proven that the Gaussian steering is monogamous and then
satisfies a CKW-type monogamy inequality \cite{Kogias2,Lami}.

Considering an arbitrary $m$-mode Gaussian state $\hat{\varrho}%
_{A_{1}...A_{m}}$ with CM $\sigma _{A_{1}...A_{m}}$, in which each party $%
A_{j}$ is a single mode-state, then, the following constraints hold \cite%
{Kogias2,Lami}
\begin{eqnarray}
\mathcal{G}^{\left( A_{1},...,A_{k-1},A_{k+1},...,A_{m}\right) \rightarrow
A_{k}}(\sigma _{A_{1}...A_{m}}) &\geqslant &\sum\limits_{j=1(j\neq k)}^{m}%
\mathcal{G}^{A_{j}\rightarrow A_{k}}(\sigma _{A_{1}...A_{m}}),\text{ \ }%
\forall k=1,...,m  \label{Mono-1} \\
\mathcal{G}^{A_{k}\rightarrow \left(
A_{1},...,A_{k-1},A_{k+1},...,A_{m}\right) }(\sigma _{A_{1}...A_{m}})
&\geqslant &\sum\limits_{j=1(j\neq k)}^{m}\mathcal{G}^{A_{k}\rightarrow
A_{j}}(\sigma _{A_{1}...A_{m}}).  \label{Mono-2}
\end{eqnarray}%
Due to the fact that a single-mode cannot be jointly steered under Gaussian
measurements, only one term $\mathcal{G}^{A_{j}\rightarrow A_{k}}$ in the
right-hand side of (\ref{Mono-1}) can be nonzero \cite%
{Reid1,Adesso-Simon,Kim-Nha}. In addition, it has been proven that one of
the constraints (\ref{Mono-1})-(\ref{Mono-2}) can be violated in the case of
more than one mode per party \cite{Lami}. So, Gaussian steering is not in
general monogamous when the common steered party is made of two or more
modes, which has been confirmed theoretically \cite{Lami} as well as
experimentally \cite{Deng1}. Keeping our focus on tripartite Gaussian states
$\hat{\varrho}_{ABC}$, the inequalities (\ref{Mono-1})-(\ref{Mono-2}) become
\cite{Kogias2}
\begin{eqnarray}
\mathcal{G}^{\left( ij\right) \rightarrow k}(\sigma _{ijk})-\mathcal{G}%
^{i\rightarrow k}(\sigma _{ijk})-\mathcal{G}^{j\rightarrow k}(\sigma _{ijk})
&\geqslant &0\text{ for }i,j,k\in \{A,B,C\},  \label{Mono-3} \\
\mathcal{G}^{k\rightarrow (ij)}(\sigma _{ijk})-\mathcal{G}^{k\rightarrow
i}(\sigma _{ijk})-\mathcal{G}^{k\rightarrow j}(\sigma _{ijk}) &\geqslant &0.
\label{Mono-4}
\end{eqnarray}%
For any bipartite ($n_{\mathcal{X}}+m_{\mathcal{Y}}$)\textit{-}mode Gaussian
state $\hat{\varrho}_{\mathcal{XY}}$ with CM $\sigma _{\mathcal{XY}}\equiv
\left( \QATOP{\mathcal{X}}{\mathcal{Z}^{\mathrm{T}}}\QATOP{\mathcal{Z}}{%
\mathcal{Y}}\right) $, the quantity $\mathcal{G}^{\mathcal{X}\rightarrow
\mathcal{Y}}$ quantifying how much--under Gaussian measurements--the party $%
\mathcal{X}$ can steer the party $\mathcal{Y}$ is defined by $\mathcal{G}^{%
\mathcal{X}\rightarrow \mathcal{Y}}(\sigma _{\mathcal{XY}}):=\max
\{0,-\sum\limits_{j:\bar{\upsilon}_{j}^{\mathcal{Y}}<1/2}\ln (\bar{\upsilon}%
_{j}^{\mathcal{Y}})\}$, where $\bar{\upsilon}_{j}^{\mathcal{Y}}$ are the
symplectic eigenvalues of the matrix $M_{\sigma }^{\mathcal{Y}}=\mathcal{Y}-%
\mathcal{Z}^{\mathrm{T}}\mathcal{XZ}$, derived from the Schur complement of $%
\mathcal{X}$ in the CM $\sigma _{\mathcal{XY}}$ \cite{Kogias1}. The
steerability $\mathcal{G}^{\mathcal{X}\rightarrow \mathcal{Y}}$ is monotone
under Gaussian local operations and classical communication, and vanishes
when the state $\sigma _{\mathcal{XY}}$ is nonsteerable by Gaussian
measurements \cite{Kogias1}. The steerability $\mathcal{G}^{\mathcal{Y}%
\rightarrow \mathcal{X}}$ can be evaluated by swapping the roles of $%
\mathcal{X}$ and $\mathcal{Y}$. As we have already mentioned above, quantum
steering is intrinsically asymmetric with respect to the role played by the
observers $\mathcal{X}$ and $\mathcal{Y}$ in a mixed state $\hat{\varrho}_{%
\mathcal{XY}}$,i.e., in general $\mathcal{G}^{\mathcal{X}\rightarrow
\mathcal{Y}}\neq \mathcal{G}^{\mathcal{Y}\rightarrow \mathcal{X}}$ \cite%
{Kogias1}. This justifies two CKW-type monogamy inequalities for steering
given by (\ref{Mono-3})-(\ref{Mono-4}). For the sake of notation, $\mathcal{G%
}^{(ij)/k}$ and $\mathcal{G}^{k/(ij)}$ will stand respectively for $\mathcal{%
G}^{\left( ij\right) \rightarrow k}-\mathcal{G}^{i\rightarrow k}-\mathcal{G}%
^{j\rightarrow k}$ and $\mathcal{G}^{k\rightarrow (ij)}-\mathcal{G}%
^{k\rightarrow i}-\mathcal{G}^{k\rightarrow j}$. So, the simultaneous
holding of the inequalities $\mathcal{G}^{(ij)/k}>0$ and $\mathcal{G}%
^{k/(ij)}>0$ $\forall $ $i,j,k\in \{A,B,C\}$, certifies that \textit{genuine
tripartite steering} is shared between the three modes $A$, $B$ and $C$.
This because $\mathcal{G}^{\left( ij\right) \rightarrow k}>0$ and $\mathcal{G%
}^{k\rightarrow \left( ij\right) }>0$ $\forall i,j,k\in \{A,B,C\}$ are
sufficient requirements to violate the corresponding biseparable model in
the tripartite mixed state $\hat{\varrho}_{ABC}$ \cite{Reid2,Kogias2}.
\begin{figure}[t]
\centerline{\includegraphics[width=0.45\columnwidth,height=5cm]{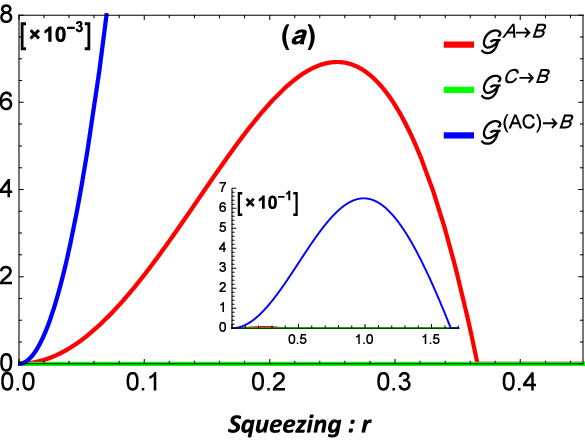}
\includegraphics[width=0.45\columnwidth,height=5cm]{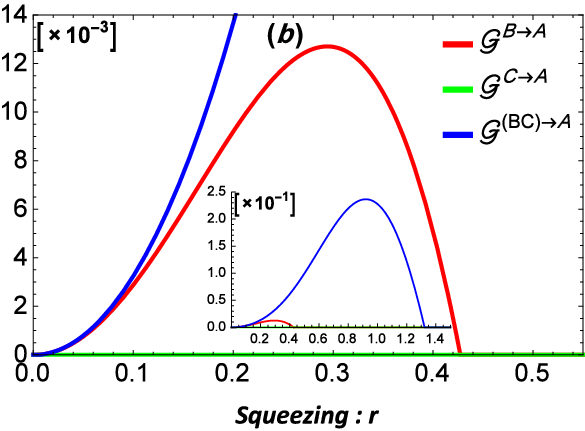}}
\centerline{\includegraphics[width=0.45\columnwidth,height=5cm]{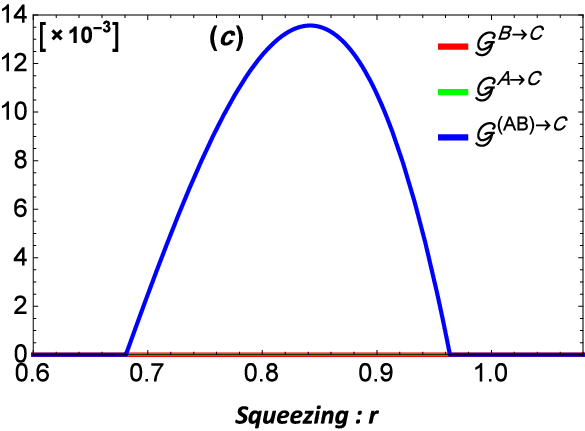}
\includegraphics[width=0.45\columnwidth,height=5cm]{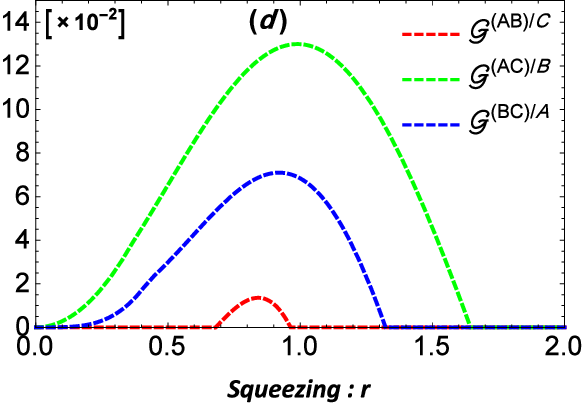}}
\caption{The Gaussian steering $\mathcal{G}^{(ij)\rightarrow k}$, $\mathcal{G%
}^{i\rightarrow k}$ and $\mathcal{G}^{j\rightarrow k}$ (panels (\textit{a}),
(\textit{b}) and (\textit{c})), and the monogamies $\mathcal{G}^{(ij)/k}=%
\mathcal{G}^{\left( ij\right) \rightarrow k}-\mathcal{G}^{i\rightarrow k}-%
\mathcal{G}^{j\rightarrow k}$ (panel (\textit{d})) for $i,j,k\in \{A,B,C\}$
vs the squeezing $r$. The case (\textit{c}) is interesting, since its
existence has been conjectured in \protect\cite{Kogias2}, and then proven to
be pertinent for practical implementation of the quantum secret sharing
protocol \protect\cite{Kogias2}.}
\label{fig2}
\end{figure}

Now, we are in position to study--\textit{under squeezing effects}--Gaussian
steering of the configurations $\mathcal{G}^{i\rightarrow j}$, $\mathcal{G}%
^{i\rightarrow (jk)}$ and $\mathcal{G}^{(ij)\rightarrow k}$ with $i,j,k\in
\{A,B,C\}$. So, we use parameters from \cite{Groblacher}. The right(left)
cavity having lengths $l_{1,2}=l=25~\mathrm{\ mm}$, decay rate $\kappa
_{_{1,2}}=\kappa =2\pi \times 215~\mathrm{KHz}$ and frequency $\omega
_{c_{1,2}}=\omega _{c}=2\pi \times 5.26\times 10^{14}~\mathrm{Hz}$ are
pumped by lasers of frequency $\omega _{L_{1,2}}=\omega _{L}=2\pi \times
2.82\times 10^{14}$ $\mathrm{Hz}$. The movable mirror has a mass $\mu =145~%
\mathrm{ng}$ and oscillating at frequency $\omega _{m}=2\pi \times 947~%
\mathrm{KHz}$. The mechanical damping rate $\gamma _{m}$ is fixed so that $%
\alpha =\frac{\gamma _{m}}{\kappa }=0.05$. Since the quantum steering is
inherently asymmetric, it may be possible to observe the steerability of a
bipartite state $\hat{\varrho}_{\mathcal{XY}}$ in one direction, but not in
the other. For this, we introduce large asymmetry in the system by using
different input laser powers,i.e., $\wp _{2}=\frac{\wp _{1}}{100}=4~\mathrm{%
mW}$.
\begin{figure}[t]
\centerline{\includegraphics[width=0.45\columnwidth,height=5cm]{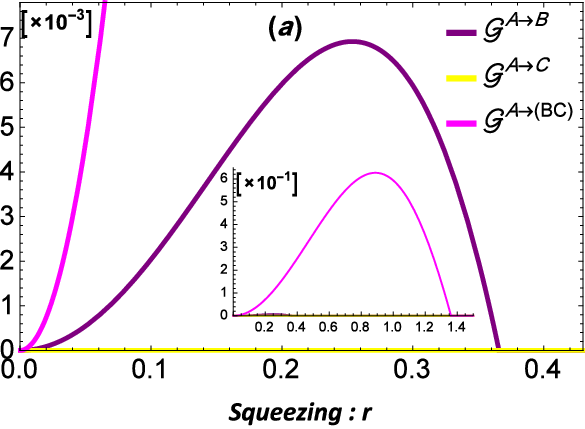}
\includegraphics[width=0.45\columnwidth,height=5cm]{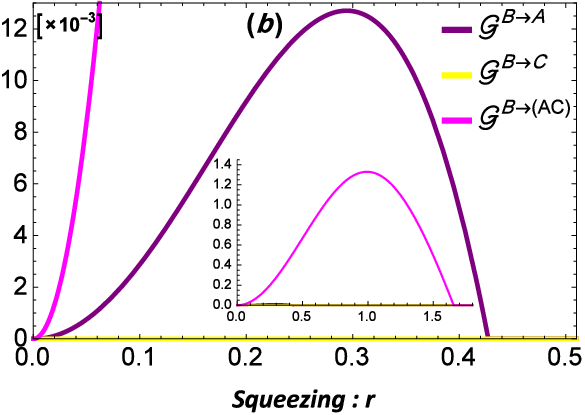}}
\centerline{\includegraphics[width=0.45\columnwidth,height=5cm]{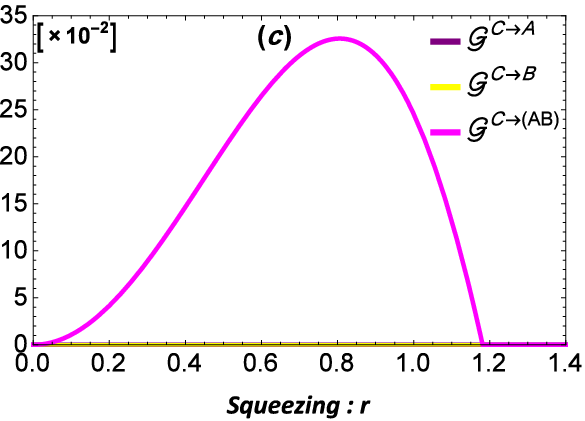}
\includegraphics[width=0.45\columnwidth,height=5cm]{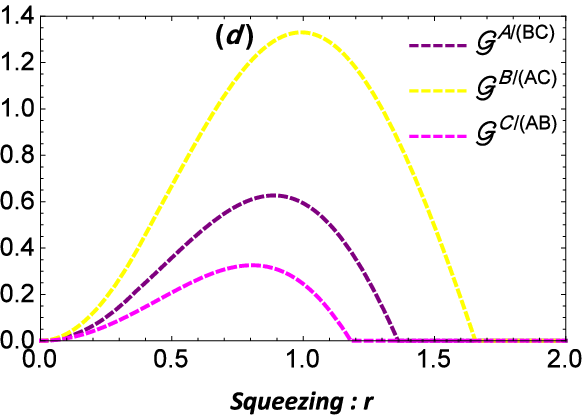}}
\caption{The Gaussian steering $\mathcal{G}^{k\rightarrow (ij)}$, $\mathcal{G%
}^{k\rightarrow i}$ and $\mathcal{G}^{k\rightarrow j}$ (panels (\textit{a}),
(\textit{b}) and (\textit{c})), and the monogamies $\mathcal{G}^{k/(ij)}=%
\mathcal{G}^{k\rightarrow \left( ij\right) }-\mathcal{G}^{k\rightarrow i}-%
\mathcal{G}^{k\rightarrow j}$ (panel (\textit{d})) for $i,j,k\in \{A,B,C\}$
vs the squeezing $r$. Panels (\textit{a}) and (\textit{b}) show that the
single-mode $k$ cannot steer simultaneously the two others single-mode $i$
and $j$. It's just a curious coincidence, since the simultaneous
steerability of an arbitrary number of single-mode by a single-mode can
happen \protect\cite{Kogias2,Lami}, as demonstrated in two different
three-mode scenarios \protect\cite{Theo1}.}
\label{fig3}
\end{figure}
Notice that the situation $\kappa /\omega _{m}\approx 0.2\ll 1$ places us
well into the resolved sideband regime \cite{Groblacher}, which justifies
the use of the RWA above. In general, it is harder to generate quantum
steering than entanglement, since steering requires strong non-separable
quantum correlations between parties and tolerates less thermal noise \cite%
{Wollmann, He and Gong}. So, we take $\bar{n}=10^{-4}$ for the mean thermal
phonons number \cite{Clerk}.

We start by discussing the behavior of the steering $\mathcal{G}^{\left(
ij\right) \rightarrow k}$, $\mathcal{G}^{i\rightarrow k}$ and $\mathcal{G}%
^{j\rightarrow k}$ for $i,j,k\in \{A,B,C\}$. Figs. \ref{fig2}$(a)$, \ref%
{fig2}$(b)$ and \ref{fig2}$(c)$ show that the single-mode $k$ cannot be
jointly steered by the two others single-mode $i$ and $j$. In other words,
if the mode $k$ could be steered by the mode $i$ ($\mathcal{G}^{i\rightarrow
k}$ $>0$), thus, it must be necessarily $\mathcal{G}^{j\rightarrow k}$ $=0$ $%
\forall i,j,k\in \{A,B,C\}$ \cite{Kogias2,Lami}. This observation can be
well understood thanks to the monogamy relation derived from the second
order moments criterion \cite{Reid1},i.e., two distinct single-mode cannot
steer a third single-mode simultaneously performing Gaussian measurements.
Notice that such scenario was generalized for the case of parties $i$ and $j$
comprising an arbitrary number of modes \cite{Adesso-Simon,Kim-Nha}, where
an experimental verification has been achieved using Gaussian cluster states
\cite{Deng1}.

As conjectured in \cite{Kogias2}, Fig. \ref{fig2}$(c)$ shows an extreme
level of steering, in which the two single-mode $A$ and $B$ cannot
individually steer the single-mode $C$,i.e., $\mathcal{G}^{A\rightarrow C}$ $%
=\mathcal{G}^{B\rightarrow C}$ $=0$, while, they can collectively,i.e., $%
\mathcal{G}^{(AB)\rightarrow C}$ $>0$. Such important scenario has been
proven to be the key resource for practical implementation of the QSS
protocol proposed recently for secure quantum communication \cite%
{Kogias2,Kogias3}. Therefore, it follows that the three-mode optomechanical
system under consideration, is an appropriate resource for the QSS protocol.

Furthermore, Fig. \ref{fig3} shows the behavior of the steering $\mathcal{G}%
^{k\rightarrow (ij)}$, $\mathcal{G}^{k\rightarrow i}$ and $\mathcal{G}%
^{k\rightarrow j}$ for $i,j,k\in \{A,B,C\}$. It is worth interesting to
remark that Fig. \ref{fig3}$(c)$ illustrates another extreme level of
steering similar to that observed in Fig. \ref{fig2}$(c)$: the single-mode $%
C $ cannot individually steer neither the mode $A$ nor $B$,i.e.,$\mathcal{G}%
^{C\rightarrow A}=\mathcal{G}^{C\rightarrow B}=0$, however, it can steer
them collectively,i.e., $\mathcal{G}^{C\rightarrow (AB)}$ $>0$. Also, Figs. %
\ref{fig3}$(a),$ \ref{fig3}$(b)$ and Fig. \ref{fig3}$(c)$ show that the
collective steerability $\mathcal{G}^{k\rightarrow (ij)}$ is always
significantly higher than the individual steerabilities $\mathcal{G}%
^{k\rightarrow i}$ and $\mathcal{G}^{k\rightarrow j}$. Strikingly, Figs. \ref%
{fig3}(\textit{a})-\ref{fig3}(\textit{b}) show that the single-mode $k,$
cannot steer simultaneously the two others single-mode $i$ and $j.$ Then, it
is important to stress that such property is not general. It is just a
\textit{curious coincidence}, because, the simultaneous steerability of an
arbitrary number of single-mode (i.e., more than one single-mode) by a
single-mode is authorized by the quantum multipartite steering theory \cite%
{Kogias2,Lami}, as has been theoretically confirmed in two different models
\cite{Theo1}. Surprisingly, as can be seen from Figs. \ref{fig2}$(a)$, \ref%
{fig2}$(b)$ and \ref{fig2}$(c)$, as well as from Figs. \ref{fig3}$(a)$, \ref%
{fig3}$(b)$ and \ref{fig3}$(c)$, we have found that no steering exist between the
interacting modes $C$ and $A$(resp. $B$), where $\mathcal{G}^{A\rightarrow
C}=\mathcal{G}^{C\rightarrow A}=0$($\mathcal{G}^{B\rightarrow C}=\mathcal{G}%
^{C\rightarrow B}=0$), despite that the mode $C$ is directly coupled via the
radiation pressure interaction with the modes $A$ and $B$. In contrast, nonzero steering
can be observed between the two optical modes $A$ and $B$, even though they
are not directly coupled. It follows that as long as the radiation pressure effect remains unable to generate quantum steering neither in the state $\hat{%
\varrho}_{AC}$ nor in $\hat{\varrho}_{BC}$, no steerability can be generated
as well in the state $\hat{\varrho}_{AB}$. Then, the steerability observed
in the different bipartitions ($A/B$, $(AB)/C $, $(AC)/B$ and $(BC)/A$) can
be regarded as a consequence of quantum fluctuations transfer from the
squeezed light to state $\hat{\varrho}_{ABC}$.

We further investigate quantitatively the behavior of the steering
monogamies (\ref{Mono-3})-(\ref{Mono-4}) against the squeezing parameter $r$%
. Quite importantly, Figs. \ref{fig2}$(d)$ and \ref{fig3}$(d)$ show that the CKW-like steering inequalities (\ref{Mono-3})-(\ref{Mono-4}%
)--imposing fundamental restrictions to the distribution of Gaussian
steering among tripartite states--are fully satisfied by the tripartite
state $\hat{\varrho}_{ABC}$. This can be interpreted as follows: in the
state $\hat{\varrho}_{ABC}$, the degree of steering exhibited--under
Gaussian measurements--in the collective configuration $\mathcal{G}^{\left(
ij\right) \rightarrow k}$ ($\mathcal{G}^{k\rightarrow \left( ij\right) }$)
can be larger than the sum of steering measured in the bimodal
configurations $\mathcal{G}^{i\rightarrow k}$ and $\mathcal{G}^{j\rightarrow
k}$($\mathcal{G}^{k\rightarrow i}$ and $\mathcal{G}^{k\rightarrow j}$) $%
\forall i,j,k\in \{A,B,C\}$. Moreover, the holding of such inequalities
means that quantum steering cannot be freely shared between the three
considered modes, which therefore has profound applications in quantum
communication \cite{Kogias2}.

\begin{figure}[t]
\centerline{\includegraphics[width=0.45\columnwidth,height=5cm]{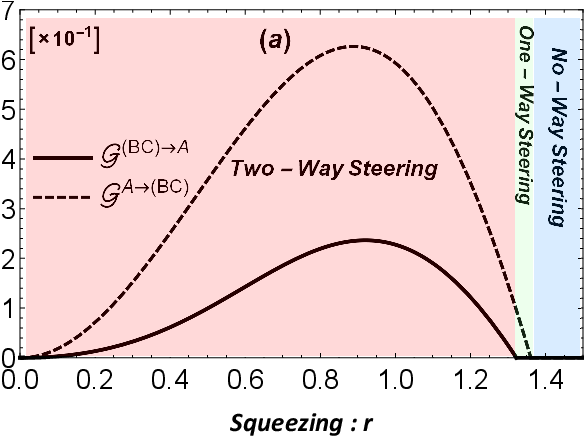}
\includegraphics[width=0.45\columnwidth,height=5cm]{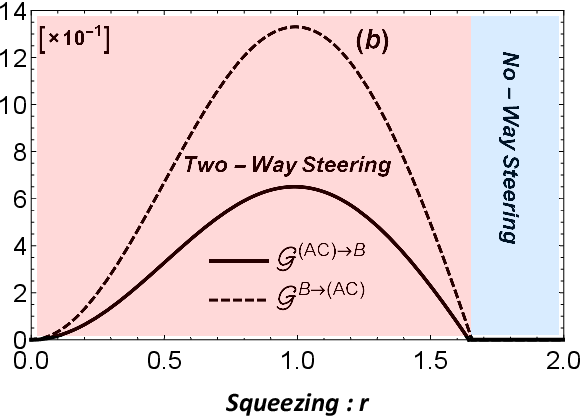}}
\centerline{\includegraphics[width=0.45\columnwidth,height=5cm]{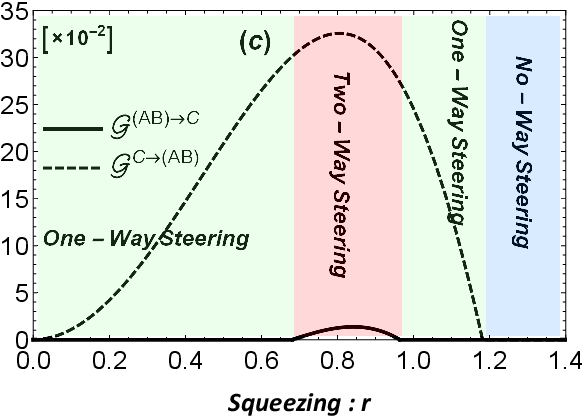}
\includegraphics[width=0.45\columnwidth,height=5cm]{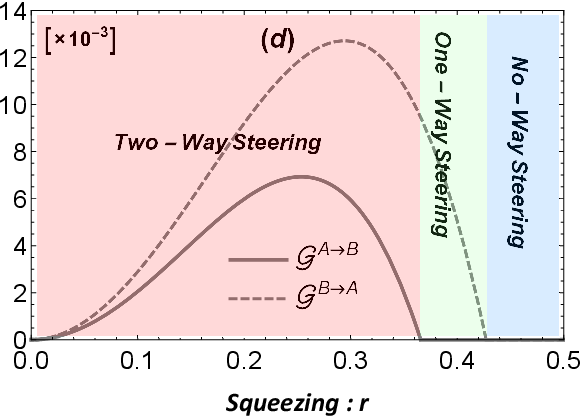}}
\caption{The Gaussian steering $\mathcal{G}^{(ij)\rightarrow k}$ (black
solid lines) vs $\mathcal{G}^{k\rightarrow (ij)}$ (black dashed lines) for $%
i,j,k\in \{A,B,C\}$ (panels (\textit{a}), (\textit{b}) and (\textit{c})),
and $\mathcal{G}^{A\rightarrow B}$ (gray solid line) vs $\mathcal{G}%
^{B\rightarrow A}$ (gray dashed line) (panel (\textit{d})) against the
squeezing $r$. The red, green and blue regions refer respectively to the
two-way, one-way and the no-way steering behaviors. Interestingly, panel (%
\textit{b}) reveals that the asymmetric aspect of a bipartite quantum state
is a necessary condition--but not sufficient--for observing one-way
steering. }
\label{fig4}
\end{figure}

An essential result shown by Figs. \ref{fig2}($d$) and \ref{fig3}($d$)
together, is that, although no direct interaction between the two optical
modes $A$ and $B$ is considered, the steering monogamies (\ref{Mono-3})-(\ref%
{Mono-4}) hold simultaneously within $0.7\leq r\leq 0.98$, which therefore
witnesses that the state $\hat{\varrho}_{ABC}$ displays \textit{genuine
tripartite steering} in this range \cite{Reid2}. This is due to the fact
that $\mathcal{G}^{\left( ij\right) \rightarrow k}>0$ and $\mathcal{G}%
^{k\rightarrow (ij)}>0$ $\forall i,j,k\in \{A,B,C\}$ are sufficient
conditions to violate the biseparable structure of the \textit{mixed state} $%
\hat{\varrho}_{ABC}$ \cite{Reid2,Kogias2}.

Now, we turn our focus into the one-way steering property as an interesting
feature of steering. In Fig. \ref{fig4}, we plot the steerability $\mathcal{G%
}^{(ij)\rightarrow k}$ versus $\mathcal{G}^{k\rightarrow (ij)}$ for $%
i,j,k\in \{A,B,C\}$, and the bi-mode steering $\mathcal{G}^{A\rightarrow B}$
versus $\mathcal{G}^{B\rightarrow A}$ as functions of the squeezing $r$. The
red, green and blue regions refer respectively to the two-way, one-way and
no-way steering behaviors. In essence, we remark that one-way steering can
be observed in the configuration (1vs2)-mode (Figs. \ref{fig4}(\textit{a})-%
\ref{fig4}(\textit{c})) as well as in the configuration (1vs1)-mode (Fig. %
\ref{fig4}(\textit{d})). Obviously, the asymmetry exhibited in the first
configuration is strongly influenced by the intermodal structure of the
considered bipartition. Indeed, Fig. \ref{fig4}(\textit{c}) shows that in
the bipartition $(AB)/C$, one-way steering is widely detected over two
ranges of the squeezing $r$. In contrast, within the bipartition $(BC)/A$
(Fig. \ref{fig4}(\textit{a})), such behavior is reached only in a narrow-band
of the parameter $r$, whereas, no one-way steering is observed within the
bipartition $(AC)/B$ (Fig. \ref{fig4}(\textit{b})). This could be deeply
related to the results presented in Figs. \ref{fig2} and \ref{fig3},i.e., no
steering exist between the mode $C$ and $A$($B$), where $\mathcal{G}%
^{A\rightarrow C}=\mathcal{G}^{C\rightarrow A}=0$($\mathcal{G}^{B\rightarrow
C}=\mathcal{G}^{C\rightarrow B}=0$). More importantly, Fig. \ref{fig4}(%
\textit{b}) shows an interesting result in which no one-way steering is
observed despite the asymmetric aspect of the bipartition $(AC)/B$. Hence,
it must be pointed out that the asymmetry introduced into a bipartite
quantum state is only a necessary--but not sufficient--condition for
reaching one-way steering. In other words, a bipartite state demonstrating
one-way steering is necessarily asymmetric, while, a bipartite state
exhibiting only two-way steering may also be an asymmetric quantum state.

It is worthy to remark from Fig. \ref{fig4}(\textit{d}) that both two-way
steering (the red region) and one-way steering (the green region) were
manifested in the bipartite state $\hat{\varrho}_{AB}$, indicating that
asymmetric nonseparable quantum correlations are shared between the modes $A$
and $B$ even though they are not directly coupled. This is necessarily due
to a mediation through the mechanical mode $C$, that can be regarded as an
auxiliary mode \cite{Paternostro2}.

Finally, we discuss the nontrivial dependence of the observed steering on
the squeezing $r$. Remarkably, Figs. \ref{fig2}-\ref{fig3} show that no
quantum steering can be detected for small values of the squeezing $r$
(including $r=0$). This is due to the fact that without squeezing ($r=0$) no
quantum fluctuations can be transferred from the squeezed light to the
system and thus, no quantum steering can be generated. Furthermore,
without--or with small--squeezing $r$, the radiation pressure
is not sufficient to generate strong quantum correlations in the considered
optomechanical system. Also, Figs. \ref{fig2}-\ref{fig3} show that all the
bipartite steering exhibited by the state $\hat{\varrho}_{ABC}$ undergo a
resonance-like behavior, i.e., the emerged steering--labelled $\mathcal{G}^{%
\mathcal{X}\rightarrow \mathcal{Y}}$--increases with increasing $r$, reaching a maximum for a given squeezing $r$ denoted $r_{0}$. In contrast,
for $r>r_{0}$, the steering $\mathcal{G}^{\mathcal{X}\rightarrow \mathcal{Y}%
} $ decreases and quickly disappears completely. Such double
effect--enhancement and degradation--of the squeezed light on the steering
can be well explained knowing that the reduced state of a two-mode squeezed
light is a thermal state with an average photons number proportional to the
squeezing degree $r$ \cite{Paternostro1}. So, for $0<r<r_{0},$ the
progressive injection of the squeezed light increases the number of photons
in the system. This enhances the optomechanical coupling via the radiation
pressure, which makes the detection of non-classical correlations--including
quantum steering--possible. While, a broadband squeezed light ($r>r_{0})$
induces strong input thermal noise, which degrades the generated
non-classical correlations.

\section{Conclusion\label{sec4}}

In an optomechanical system fed by squeezed light, we have studied the
distribution of EPR-steering over an asymmetric three-mode Gaussian state $%
\hat{\varrho}_{ABC}$. The stationary covariance matrix describing the state $%
\hat{\varrho}_{ABC}$ is obtained in the resolved-sideband regime. Using the
measure proposed in \cite{Kogias1}, we showed that a single-mode $k$ cannot
be jointly steered by the two other modes $i$ and $j$ for $i,j,k\in
\{A,B,C\}$. We have also provided theoretical confirmation for two types of
Coffman-Kundu-Wootters-like steering monogamy relations in the tripartite
state $\hat{\varrho}_{ABC}$, which thus bound the distribution of steering
among the three modes $A,B$ and $C$. Important result is that, significant
\textit{genuine tripartite Gaussian steering }has been displayed by the
state $\hat{\varrho}_{ABC}$, which witnesses the presence of strong quantum
correlations between the three considered modes. Interestingly enough, an
extreme level of steering, where its existence has been conjectured in \cite%
{Kogias2}, is straightforwardly reached, i.e., as illustrated in Fig. \ref%
{fig2}(\textit{c}), the modes $A$ and $B$ cannot steer individually the mode
$C$ ($\mathcal{G}^{A\rightarrow C}$ $=\mathcal{G}^{B\rightarrow C}$ $=0$),
while, they can collectively ($\mathcal{G}^{(AB)\rightarrow C}$ $>0$). Such
scenario is considered as the key ingredient allowing practical
implementation of the quantum secret sharing protocol proposed for secure
quantum communication \cite{Kogias2,Kogias3}. Moreover, Gaussian one-way
steering has been observed in the individual configuration (1vs1)-mode as
well as in the collective one (1vs2)-mode. As another fundamental result, we
have found that, the asymmetry introduced into a bipartite state $\hat{%
\varrho}_{\mathcal{XY}}$ is just a necessary condition--but not
sufficient--for observing the one-way steering property (see Fig. \ref{fig4}(%
\textit{b})).

This work constitutes a first step towards a depth understanding of Gaussian
EPR-steering and its monogamy in asymmetric three-mode states. This may
provide a useful resource for the quantum secret sharing protocol as well as
for one-way quantum tasks.


\begin{thebibliography}{99}
\bibitem{Bell} J. S. Bell, Physics \textbf{1}, 195 (1964).

\bibitem{Svetlichny} G. Svetlichny, Phys. Rev. D \textbf{35}, 3066 (1987).

\bibitem{Reid2} Q. Y. He and M. D. Reid, Phys. Rev. Lett. \textbf{111},
250403 (2013).

\bibitem{Wiseman} H. M. Wiseman, S. J. Jones, and A. C. Doherty, Phys. Rev.
Lett. \textbf{98}, 140402 (2007); S. J. Jones, H. M. Wiseman, and A. C.
Doherty, Phys. Rev. A \textbf{76}, 052116 (2007).

\bibitem{entanglement} R. Horodecki, P. Horodecki, M. Horodecki, and K.
Horodecki, Rev. Mod. Phys. \textbf{81}, 865 (2009).

\bibitem{discord} G. Adesso and A. Datta, Phys. Rev. Lett. \textbf{105},
030501 (2010); P. Giorda and M. G. A. Paris, Phys. Rev. Lett. \textbf{105},
020503 (2010).

\bibitem{Chuang} M. A. Nielsen and I. L. Chuang, Quantum Computation and
Quantum Information (Cambridge University Press, Cambridge, England, 2000).

\bibitem{CKW} V. Coffman, J. Kundu, and W. K. Wootters, Phys. Rev. A \textbf{%
61}, 052306 (2000).

\bibitem{Adesso3} C. Lancien, S. Di Martino, M. Huber, M. Piani, G. Adesso,
and A. Winter, Phys. Rev. Lett. \textbf{117}, 060501 (2016).

\bibitem{Adesso1} G. Adesso, D. Girolami, and A. Serafini, Phys. Rev. Lett.
\textbf{109}, 190502 (2012).

\bibitem{Reid1} M. D. Reid, Phys. Rev. A \textbf{88}, 062108 (2013).

\bibitem{Kogias1} I. Kogias, A. R. Lee, S. Ragy, and G. Adesso, Phys. Rev.
Lett. \textbf{114}, 060403 (2015).

\bibitem{Kogias2} Y. Xiang, I. Kogias, G. Adesso and Q. He, Phys. Rev. A
\textbf{95}, 010101(R) (2017).

\bibitem{Lami} L. Lami, C. Hirche, G. Adesso, and A. Winter, Phys. Rev.
Lett. \textbf{117}, 220502 (2016).

\bibitem{Schrodinger} E. Schr\"{o}dinger, Math. Proc. Cambridge Philos. Soc.
\textbf{32}, 446 (1936); E. Schr\"{o}dinger, Math. Proc. Cambridge Philos.
Soc. \textbf{31}, 555 (1935).

\bibitem{Exp} S. Armstrong, M. Wang, R. Y. Teh, Q. H. Gong, Q. Y. He, J.
Janousek, H. A. Bachor, M. D. Reid, P. K. Lam, Nat. Phys. \textbf{11} 167
(2015).

\bibitem{Handchen} V. H\"{a}ndchen, T. Eberle, S. Steinlechner, A.
Samblowski, T. Franz, R. F. Werner, and R. Schnabel, Nature Photon \textbf{6}%
, 598 (2012).

\bibitem{Wollmann} S. Wollmann, N. Walk, A. J. Bennet, H. M. Wiseman and G.
J. Pryde, Phys. Rev. Lett. \textbf{116}, 160403 (2016).

\bibitem{Farzad} N. Tischler, F. Ghafari, T. J. Baker, S. Slussarenko, R. B.
Patel, M. M. Weston, S. Wollmann, L. K. Shalm, V. B. Verma, S. W. Nam, H. C.
Nguyen, H. M. Wiseman, and G. J. Pryde, Phys. Rev. Lett. \textbf{121},
100401 (2018).

\bibitem{Bowles} J. Bowles, T. V\'{e}rtesi, M. T. Quintino, and N. Brunner,
Phys. Rev. Lett. \textbf{112}, 200402 (2014).

\bibitem{Branciard} C. Branciard, E. G. Cavalcanti, S. P. Walborn, V.
Scarani, and H. M. Wiseman, Phys. Rev. A \textbf{85}, 010301(R) (2012).

\bibitem{Skrzypczyk} P. Skrzypczyk, and D. Cavalcanti, Phys. Rev. Lett.
\textbf{120}, 260401 (2018).

\bibitem{Goswami} S. Goswami, B. Bhattacharya, D. Das, S. Sasmal, C.
Jebaratnam, and A. S. Majumdar, Phys. Rev. A \textbf{98}, 022311 (2018).

\bibitem{Piani} M. Piani, and J. Watrous, Phys. Rev. Lett. \textbf{114},
060404 (2015).

\bibitem{zarate} Q. He, L. Rosales-Zarate, G. Adesso, and M. D. Reid, Phys.
Rev. Lett. \textbf{115}, 180502 (2015).

\bibitem{Kogias3} I. Kogias, Y. Xiang, Q. He, and G. Adesso, Phys. Rev. A
\textbf{95}, 012315 (2017).

\bibitem{Peres} T. Moroder, O. Gittsovich, M. Huber, and O. G\"{u}hne, Phys.
Rev. Lett. \textbf{113}, 050404 (2014); S. Yu and C. H. Oh, Phys. Rev. A
\textbf{95}, 032111 (2017).

\bibitem{Theo1} Q. He and Z. Ficek, Phys. Rev. A \textbf{89}, 022332 (2014);
J. Wang, J. Jing, and H. Fan, Ann. Phys. (Berlin) \textbf{530}, 1700261
(2018).

\bibitem{Theo2} M. Wang, Q. Gong, Z. Ficek, and Q. He, Phys. Rev. A \textbf{%
90}, 023801 (2014); M. Wang, Z. Qin, Y. Wang, and X. Su, Phys. Rev. A
\textbf{96}, 022307 (2017).

\bibitem{Deng1} X. W. Deng, Y. Xiang, C. X. Tian, G. Adesso, Q. Y. He, Q. H.
Gong, X. L. Su, C. D. Xie, K. C. Peng, Phys. Rev. Lett. \textbf{118} 230501
(2017).

\bibitem{Deng2} X. Deng, C. Tian, M. Wang, Z. Qin, X. Su, Opt. Commun.
\textbf{421} 14 (2018).

\bibitem{Marquardt} M. Aspelmeyer, T. J. Kippenberg, and F. Marquardt, Rev.
Mod. Phys. \textbf{86}, 1391 (2014).

\bibitem{cooling} I. Wilson-Rae, N. Nooshi, W. Zwerger, and T. J.
Kippenberg, Phys. Rev. Lett. \textbf{99}, 093901 (2007); J. D. Teufel, T.
Donner, D. Li, J. W. Harlow, M. S. Allman, K. Cicak, A. J. Sirois, J. D.
Whittaker, K. W. Lehnert and R. W. Simmonds, Nature (London) \textbf{475},
359 (2011).

\bibitem{Qtransfer} T. A. Palomaki, J. W. Harlow, J. D. Teufel, R. W.
Simmonds, and K. W. Lehnert, Nature (London) \textbf{495}, 210 (2013); J. El
Qars, M. Daoud, and R. A. Laamara, J. Mod. Opt. \textbf{65}, 1584 (2018).

\bibitem{Entanglements} G. De Chiara, M. Paternostro, and G. M. Palma, Phys.
Rev. A \textbf{83}, 052324 (2011); J. El Qars, M. Daoud, and R. A. Laamara,
Int. J. Mod. Phys. B \textbf{30}, 1650134 (2016).

\bibitem{Paternostro2} M. Paternostro, D. Vitali, S. Gigan, M. S. Kim, C.
Brukner, J. Eisert, and M. Aspelmeyer, Phys. Rev. Lett. \textbf{99}, 250401
(2007).

\bibitem{steering} J. E. Qars, M. Daoud and R. A. Laamara, Phys. Rev. A
\textbf{98}, 042115 (2018); J. E. Qars, M. Daoud and R. A. Laamara, Eur.
Phys. J. D \textbf{71}, 122 (2017); J. Li and S. -Y. Zhu, Phys. Rev. A
\textbf{96}, 062115 (2017).

\bibitem{Marinkovic} I. Marinkovi\'{c}, A. Wallucks, R. Riedinger, S. Hong,
M. Aspelmeyer, and S. Gr\"{o}blacher, Phys. Rev. Lett. \textbf{121}, 220404
(2018).

\bibitem{YXiang} Y. Xiang, F. X. Sun, M. Wang, Q. H. Gong, and Q. Y. He,
Opt. express \textbf{23}, 30104 (2015).

\bibitem{Law} C. K. Law, Phys. Rev. A \textbf{51}, 2537 (1995).

\bibitem{Genes} C. Genes, A. Mari, D. Vitali, and P. Tombesi, Adv. At. Mol.
Opt. Phys. \textbf{57}, 33 (2009).

\bibitem{Benguria} R. Benguria and M. Kac, Phys. Rev. Lett. \textbf{46}, 1
(1981).

\bibitem{Paternostro1} L. Mazzola and M. Paternostro, Phys. Rev. A \textbf{83%
}, 062335 (2011).

\bibitem{Parkins} A. S. Parkins and H. J. Kimble, J. Opt. B \textbf{1}, 496
(1999).

\bibitem{Vitali} D. Vitali, S. Gigan, A. Ferreira, H. R. B\"{o}hm, P.
Tombesi, A. Guerreiro, V. Vedra, A. Zeilinger, and M. Aspelmeyer, Phys. Rev.
Lett. \textbf{98}, 030405 (2007).

\bibitem{DWang} Y.-D. Wang and A. A. Clerk, Phys. Rev. Lett. \textbf{108},
153603 (2012).

\bibitem{Clerk} Y. -D. Wang and A. A. Clerk, Phys. Rev. Lett. \textbf{110},
253601 (2013).

\bibitem{Purdy} T. P. Purdy, P. -L. Yu, R. W. Peterson, N. S. Kampel, and C.
A. Regal, Phys. Rev. X \textbf{3}, 031012 (2013).

\bibitem{Parks} P. C. Parks and V. Hahn, Stability Theory (Prentice-Hall,
New York, 1993).

\bibitem{Dhar} H. S. Dhar, A. K. Pal, D. Rakshit, A. Sen(De), and U. Sen,
Lectures on General Quantum Correlations and their Applications (Springer
International, New York, 2017), pp. 23--64.

\bibitem{Illuminati} G. Adesso, A. Serafini, and F. Illuminati, Phys. Rev. A
\textbf{73}, 032345 (2006).

\bibitem{Adesso-Simon} G. Adesso and R. Simon, J. Phys. A \textbf{49},
34LT02 (2016).

\bibitem{Kim-Nha} S. -W. Ji, M. S. Kim, and H. Nha, J. Phys. A: Math. Theor.
\textbf{48}, 135301 (2015).

\bibitem{Groblacher} S. Gr\"{o}blacher, K. Hammerer, M. R. Vanner, and M.
Aspelmeyer, Nature (London) \textbf{460}, 724 (2009); E. A. Sete, H. Eleuch,
and C. H. R Ooi, J. Opt. Soc. Am. B \textbf{31,} 2821 (2014).

\bibitem{He and Gong} Q. Y. He, Q. H. Gong, and M. D. Reid, Phys. Rev. Lett.
\textbf{114}, 060402 (2015).
\end{thebibliography}
\end{document}